\documentclass{optica-article}

\journal{oe}

\usepackage{diagbox}
\usepackage{booktabs}
\usepackage{listings}
\usepackage{verbatim}
\usepackage{lineno}
\usepackage{caption}
\usepackage{array}%
\usepackage{stfloats}
\usepackage{float}
\usepackage{hyperref}
\hypersetup
{
    colorlinks=true,
    urlcolor=urlblue,
    linkcolor=urlblue,
    citecolor=urlblue,
    breaklinks=true,
}
\begin{document}
\title{Physics-informed neural network for nonlinear dynamics of self-trapped necklace beams}

\author{Dongshuai Liu,\authormark{1} Wen Zhang,\authormark{2} Yanxia Gao,\authormark{3} Dianyuan Fan,\authormark{1} Boris A. Malomed, \authormark{4,5} and Lifu Zhang\authormark{1,*}}

\address{
    \authormark{1}International Collaborative Laboratory of 2D Materials for Optoelectronic Science \& Technology, Institute of Microscale Optoelectronic, Shenzhen University, Shenzhen 518060, China\\
    \authormark{2}School of Mathematics and Statistics, Fuyang Normal University, Fuyang 236037, China\\
    \authormark{3}School of Physics and Optoelectronic Engineering, Shenzhen University, Shenzhen 518060, China\\
    \authormark{4}Department of Physical Electronics, School of Electrical Engineering, Faculty of Engineering, Tel Aviv University, Tel Aviv 69978, Israel\\
    \authormark{5}Instituto de Alta Investigaci\'{o}n, Universidad de Tarapac\'{a}, Casilla 7D, Arica, Chile
}

\email{\authormark{*}zhanglifu@szu.edu.cn} %% email address is required

\begin{abstract}
  A physics-informed neural network (PINN) is used to produce a variety of self-trapped necklace solutions of the (2+1)-dimensional nonlinear Schr\"{o}dinger/Gross-Pitaevskii equation. We elaborate the analysis for the existence and evolution of necklace patterns with integer, half-integer, and fractional reduced orbital angular momenta by means of PINN. The patterns exhibit phenomena similar to rotation of rigid bodies and centrifugal force. Even though the necklaces slowly expand (or shrink), they preserve their structure in the course of the quasi-stable propagation over several diffraction lengths, which is completely different from the ordinary fast diffraction-dominated dynamics. By comparing different ingredients, including the training time, loss value and $\mathbb{L}_{2}$ error, PINN accurately predicts specific nonlinear dynamical properties of the evolving necklace patterns. Furthermore, we perform the data-driven discovery of parameters for both clean and perturbed training data, adding $1\%$ random noise in the latter case. The results reveal that PINN not only effectively emulates the solution of partial differential equations, but also offers applications for predicting the nonlinear dynamics of physically relevant types of patterns.
\end{abstract}

\section{Introduction}
\noindent Dynamics of nonlinear optical and matter waves (in particular, solitons), governed by the cubic nonlinear Schr\"{o}dinger equations (NLSEs), is a topic of fundamental interest \cite{kivshar2003optical,malomed_soliton_2022}. Except for the classical case of the one-dimensional (1D) cubic NLSE \cite{shabat1972exact}, such nonlinear partial differential equations (PDEs) generally have no exact analytical solutions. Although diverse numerical methods, such as the finite-difference algorithm and Newton's iterations, as well as the spectral method, have been used to study the dynamics of optical solitons, obstacles still impede solving physical problems -- in particular, in the case of incomplete initial and boundary conditions. In such cases, the numerical methods may be prohibitively difficult, or even impossible for solving PDEs with limited data. Therefore, developing more efficient techniques for solving NLSEs and thus exploring the propagation dynamics of optical and matter-wave solitons remains a relevant objective.

With the explosive development of computational capabilities and artificial-intelligence technology, data analysis and neural network-based deep learning have been rapidly advancing in diverse areas, including various nonlinear dynamics systems \cite{jiang_physics-informed_2022, zhu_predicting_2022, jaganathan_data-driven_2023, wu_predicting_2021, jiang_predicting_2024}, computational fluid dynamics \cite{zhang_cpinns_2023, li2021deep, zhang_robust_2023}, automatic speech recognition \cite{purwins_deep_2019}, unmanned driving \cite{lv_driving-style-based_2019}, \textit{etc}. Because the deep learning is primarily used as a black-box tool based on the data-driven concept without any prior knowledge, its performance is heavily dependent on the quantity and quality of data. Moreover, purely data-driven networks are often trained without considering the underlying physics. Nevertheless, the prior knowledge of physics is indispensable for full understanding of the so predicted nonlinear dynamics.

To develop an appropriate approximation for the physical model, the physical-informed neural networks (PINN) method has been proposed, which embeds the underlying physical equations and respective constraint conditions of the target problem into the neural network \cite{raissi_physics-informed_2019}. The governing equations and physical constraints, including the initial and boundary conditions, constitute the network loss function and serve as regularization mechanisms. The PINN method transforms the problem of solving the nonlinear PDE into the optimization problem for the loss function. Through this scenario, the
network integrates prior knowledge of the setting's mathematics and physics throughout the whole training process. For handling the constraint conditions, an automatic differentiation technique is used to calculate
differential terms of PDE, without the need for the respective precise data. As a result, the model is much less data-dependent. Compared to the traditional numerical calculation method, the computational complexity of PINN with a stable structure does not seem much more challenging. Another advantage of the PINN method is that it can solve the inverse problem, restoring parameters of the underlying physical model from observable dynamical data. For high-dimensional settings, the mesh-free techniques of PINN can eliminate the dimensional catastrophe \cite{yin_multi-view_2023}, to some extent (this \textquotedblleft catastrophe\textquotedblright implies that the amount of computation of vectors exponentially grows with the increase of the dimension). Thus, PINN combines the advantages of numerical and data-driven methods, while overcoming their limitations.

In nonlinear optics, the creation of various types of spatial optical solitons is chiefly provided by self-trapping of optical fields in the free space \cite{soljacic_self-trapping_1998, davydova_stable_2004,
grow_collapse_2007}, as well as in external potentials \cite{dong_vortex_2022, liu_matter-wave_2023, dong_stable_2023, liu_higher-charged_2023, liu_gap_2023, dong_rotating_2022, liu_multi-stable_2023, dong_fractional_2023, liu_transformation_2023}. Presently, the evolution dynamics of spatial optical solitons is mainly predicted by the PINN method realized in the 1D geometry \cite{li_solving_2021, meiyazhagan_data_2022, yin_dynamic_2023, wang_data-driven_2023, xu_prediction_2023}. In particular, a variety of solitons were predicted under the action of $\mathcal{PT}$-symmetric potentials \cite{zhong_data-driven_2022, zhong_data-driven_2023}. The vector solitons of coupled nonlocal nonlinear Schr\"{o}dinger equation were also predicted via improved PINN algorithm \cite{qiu_data-driven_2024, qin_wpinn_2023}. In nonlinear fractional systems with saturable nonlinearity, the evolution of multipole soliton families was studied by dint of PINN and the numerical power-conserving square-operator method \cite{bo_prediction_2023}. Furthermore, fundamental solitons were investigated in 2D systems with a $\mathcal{PT}$-symmetric optical lattice, also by dint of the PINN method \cite{wang_data-driven_2023}. However, despite the great progress in predicting the propagation dynamics of solitons by PINN techniques, the prediction of dynamics of complex patterns in the free space has not been reported in the framework of high-dimensional NLSEs. In particular, a natural question is whether the PINN method can predict the evolution of necklace-like patterns, as one of the basic types of complex structures, in the high-dimensional free space.

In this study, we elaborate the PINN method to predict the dynamics of optical necklaces composed of different numbers of bright spots. The existence, evolution and robustness of the necklaces carrying integer and fractional reduced orbital angular momentum are investigated by means of PINN. In particular, effects of the centrifugal force are demonstrated. Comparing different ingredients, including the training time, loss value and $\mathbb{L}_{2}$ error, we find that, using even fewer network layers, one can predict the necklace patterns quite accurately. The results reveal better performance of the PINN method in solving forward problems and predicting soliton dynamics. We also derive reverse deduction of coefficients of the underlying physical model.

The following presentation is organized as follows. In Sec. II, we provide a detailed introduction to the PINN method and physical model. In Sec. III, we analyze the necklace complexes carrying integer, semi-integer, and fractional values of the reduced orbital angular momentum. In Sec. IV, the inverse problem for the discovery of parameters is demonstrated. The paper is concluded by Sec. V.

%%%%%%%%%%%%%%%%%%%%%%%%%%%%%%%%%%%%%%%%%%%%%%%%%%%%%%%%%%%%%%%%%%%%%%%%%%%%%%%%%%%%%%%%

\section{The methodology and physical model}\label{sec2}

The propagation dynamics of paraxial beams in nonlinear optics is modeled by the 2D normalized NLSE \cite{kivshar2003optical}. As the NLSE contains differential terms, the numerical methods require a larger number of iterations with smaller stepsizes, to improve the accuracy. It is well known that neural networks can approximate any function by means of general function approximators \cite{hornik_multilayer_1989}. As a result, the networks can be used directly to deal with nonlinear problems, relaxing limitations such as presets, linearization, or time stepping. As concerns the accuracy, PINN makes the full use of the prior knowledge of basic mathematics and physics of the system, to provide an efficient method for solving complex nonlinear dynamical equations. Below, we elaborate on the PINN theory and explain its modeling principles in the application to nonlinear dynamics.

%%%%%%%%%%%%%%%%%%%%%%%%%%%%%%%%%%%%%%%%%%%%%%%%%%%%%%%%%%%%%%%%%%%

\subsection{The PINN methodology}

The PINN framework needs to be specifically designed for a given physical system. The general form of the (2+1)D PDE is presented in the following form, including the governing equation along with initial and boundary conditions:
\begin{equation}
\psi_{t}+\mathcal{N}(\delta,\psi,\psi_{x},\psi_{y},\psi_{xx},\psi_{yy},%
    \cdots)=0, x \in \Omega_1, \; y \in \Omega_2, \; t \in [t_1, t_2],
    \label{eq:refname1}
\end{equation}
where $\psi (t,x,y)$ is the solution of the PDE, $\delta $ denotes the set of the model's parameters, and $\mathcal{N}[\delta ,\cdots ]$ is a combination of linear and nonlinear operators. Subscripts added to $\psi $ denote the partial derivatives, $[t_{1},t_{2}]$ and $\Omega _{1,2}$ being the respective temporal and spatial computational domains. Equation (\ref{eq:refname1}) serves as the set of the underlying physical constraints, thus forming a multi-layer feedforward neural network $\widehat{\psi }(t,x,y) $ and PINN $f(t,x,y)$. They share parameters with each other, including weights, deviations and scaling factors. In this context, the temporal and spatial derivatives and nonlinear terms need to be accurately calculated. In particular, the derivatives can be obtained by dint of the automatic differentiation \cite{raissi_physics-informed_2019}.

With the help of the limited-memory Broydem-Fletcher-Goldfarb-Shanno (L-BFGS) \cite{liu_limited_1989} and Adam \cite{kingma_adam_2014} optimization techniques, the neural-network's shared parameters can be found
by minimizing the mean square error (MSE) loss caused by the feedforward network $\widehat{\psi }(t,x,y)$ and PINN $f(t,x,y)$. The loss function $\mathcal{L}$ has the following form:
\begin{equation}
\begin{split}
\mathcal{L} =\mathcal{L}_{\psi }+\mathcal{L}_{f} = & \frac{1}{n_{\psi }}\sum_{j=1}^{n_{\psi }}|\psi (t_{\psi }^{j},x_{\psi
}^{j},y_{\psi }^{j})-\psi ^{j}|^{2} \\
& \quad +\frac{1}{n_{f}}\sum_{j=1}^{n_{f}}|\psi
_{t}(t_{f}^{j},x_{f}^{j},y_{f}^{j})+\mathcal{N}(\delta ,\cdots )|^{2},
\end{split}
\label{eq:refname2}
\end{equation}

\noindent where $\mathcal{L}_{\psi }$ corresponds to the initial and boundary data, and $\mathcal{L}_{f}$ penalizes residuals of the governing equation (\ref{eq:refname1}) at a finite set of collocation points, while $n_{\psi }$ and $n_{f}$ are the numbers of auxiliary coordinates needed to calculate each MSE term. The initial and boundary training data on $\psi (t,x,y)$ can be obtained from $\{t_{\psi }^{j},x_{\psi }^{j},y_{\psi }^{j}\}_{j=1}^{n_{\psi }}$, while the collocation points of $f(t,x,y)$ are specified by $\{t_{f}^{j},x_{f}^{j},y_{f}^{j}\}_{j=1}^{n_{f}}$. Note that the boundary conditions given in Eq. (\ref{eq:refname2}) are of the Dirichlet type. When the PDE is to be solved with Neumann's or mixed boundary conditions, the boundary derivatives should also be included. The PINN method can accurately capture the intricate nonlinear dynamics of the PDE after several iterations \cite{raissi_physics-informed_2019}.

\begin{figure}[tbph]
\centering
\includegraphics[width=0.4\textwidth]{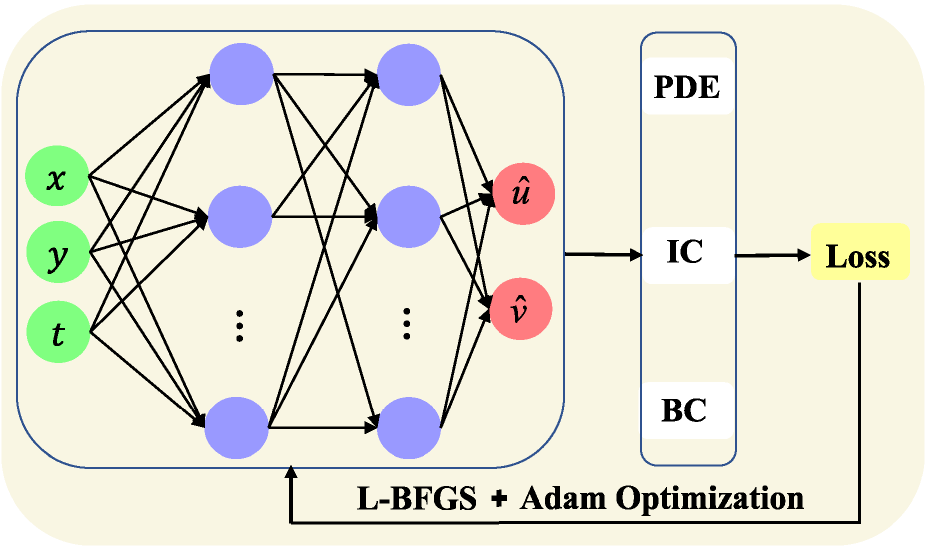}
% Here is how to import EPS art
% \captionsetup{labelformat=empty}
\caption{The main framework of the physics-informed neural network scheme. Sets $\left\{ x,y,t\right\} $ and $\left\{ \hat{u},\hat{v}\right\} $ represent the input and output variables, respectively. IC and BC stand for the initial conditions and boundary conditions, respectively.}
\label{fig1}
\end{figure}

It is necessary to build PINN composed of multiple layers of the deep neural network for solving the PDE with the initial and boundary conditions, as defined in Eq. (\ref{eq:refname1}). The auxiliary coordinates, i.e., the spatiotemporal domains and predicted solution $\widehat{\psi }(t,x,y)$ constitute the input and output of PINN, respectively. More specifically, the output of the
neural network is given as the composition of affine transformations and the nonlinear activation function: $\psi_\mathcal{X}(\phi) =(\mathcal{L}_j \; o \; \sigma \; o \; \mathcal{L}_{j-1}\; o \; \cdots \; o \; \sigma \; o \;\mathcal{L}_1)(\phi^0)$, where $\mathcal{X}$ is the full set of the trainable parameters, and $o$ stands for the composition operator. In the neural network with $j$ neurons, $\mathcal{L}_{k}(\phi ^{k-1})=W^{k}\phi ^{k-1}+b^{k}$ is the affine transformation. Here network weights $W^{k}$ and bias term $b^{k}$ are chosen from independent and identically distributed sampling. Further, the neuron activation function $\sigma $ acts as a vector activator in each hidden layer, where every neuron has its own slope as an ingredient of the activation function \cite{jagtap_locally_2020}. It is defined as $\sigma (\mathcal{L}_{k}(\phi ^{k-1})_{j}), \quad j, k=1,2, \cdots$. The nonlinear activation function $\sigma (\cdot)$ is applied to each component of the transformed vector before sending it as an input to the next layer. Common nonlinear activation functions include $\mathrm{Sigmoid}$, $\mathrm{Relu}$ (Rectified Linear Unit), $\mathrm{Selu}$ (Scaled Exponential Linear Unit), \textit{etc}. In this paper, the hyperbolic tangent $\mathrm{tanh}$ is used as the nonlinear activation function in the following. The resulting optimization problem aims at finding the minimum value of the loss function by optimizing the activation slope as well as the weights and biases. To solve the PDE with the initial and boundary conditions given in Eq.~(\ref{eq:refname1}), a PINN with multilayer deep neural network is constructed to approximate the solution, as shown in Fig.~\ref{fig1}. The main framework of the PINN scheme consists of an input layer, several hidden layers with a certain number of neurons per a layer, and an output layer. The output variables $\hat{u}$ and $\hat{v}$ must satisfy the initial conditions, boundary conditions, and governing equations (PDEs). If the initial condition is not satisfied, it needs to be corrected instantaneously, so that the next moment can be predicted. Thus the loss function is made up of three parts that are involved in the optimization process of the neural network.

%%%%%%%%%%%%%%%%%%%%%%%%%%%%%%%%%%%%%%%%%%%%%%%%%%%%%%%%%%%%%%%%%%%%%%%

\subsection{The theoretical model}

It is well known that solitons produced by the (2+1)D cubic self-focusing NLSE are unstable against the critical collapse, i.e., catastrophic self-compression or expansion, which occurs when the input's norm takes place, severally, above or below a critical value \cite{soljacic_self-trapping_2000}. The separatrix between the compression and expansion is represented by fundamental (zero-vorticity) \textit{Townes solitons} (TSs), whose norm is exactly equal to the critical value \cite{chiao_self-trapping_1964,fibich2015nonlinear}. Accordingly, the TS family is degenerate, in the sense that all the TSs have the same (critical) norm, irrespective of their propagation constant, and they are unstable against the onset of the collapse \cite{fibich2015nonlinear}. Nevertheless, because the initial growth of the instability is slow (subexponential), TSs have been observed experimentally, under carefully controlled conditions, in binary Bose-Einstein condensates \cite{bakkali-hassani_realization_2021,chen_observation_2021}.

Remarkably, self-trapped necklace-shaped beams, built as circular chains of quasi-TS elements, may demonstrate effective robustness as a whole entity, while their constituent elements are subject to collapsing \cite{soljacic_integer_2001}. It is therefore natural to employ such robust patterns to verify the performance and validity of the PINN techniques. To this end, we consider a necklace beam as a spatial-domain optical soliton complex, carrying angular momentum, which propagates along the $z$ axis in the self-focusing bulk Kerr medium. In terms of the Cartesian coordinates $x,y$ and $z$, the scaled (2+1)D NLSE for the slowly varying complex envelope of the electromagnetic field, $\psi (z,x,y)$, along with the initial and boundary conditions (ICs and BCs), is written as
\begin{equation}
\begin{split}
& i\frac{\partial \psi }{\partial {z}}+\frac{1}{2}\left( \frac{\partial
^{2}\psi }{\partial x^{2}}+\frac{\partial ^{2}\psi }{\partial y^{2}}\right)
+|\psi |^{2}\psi =0, \\
& x\in (x_{1},x_{2}),\;y\in (y_{1},y_{2}),\;z\in (0,Z), \\
& \psi (x,y,0)=\psi _{0}(x,y),\;x\in (x_{1},x_{2}),\;y\in (x_{1},x_{2}), \\
& \psi (x_{1},y,z)=\psi (x_{2},y,z),\;y\in (y_{1},y_{2}),\;z\in (0,Z), \\
& \psi (x,y_{1},z)=\psi (x,y_{2},z),\;x\in (x_{1},x_{2}),\;z\in (0,Z).
\end{split}
\label{eq:refname4}
\end{equation}%
The beam is characterized by its power (alias the above-mentioned norm), $P=\iint |\psi |^{2}dxdy$, and $z$-component of the angular momentum,
\begin{equation}
L\mathbf{\hat{z}}={i}/{2}\iint \mathbf{r}\times \{\psi \nabla \psi
^{\ast }-\psi ^{\ast }\nabla \psi \}dxdy,  \label{L}
\end{equation}%
with $\mathbf{r}=\left\{ x,y\right\} $ and $\ast $ standing for the complex conjugate. In terms of the polar coordinates $\left( r,\theta \right) $, the expression for the angular momentum is simpler,
\begin{equation}
L=\int_{0}^{2\pi }d\theta \int_{0}^{\infty }rdr~\mathrm{Im}\left( \psi
^{\ast }\frac{\partial \psi }{\partial \theta }\right) .  \label{L2}
\end{equation}%
When considering the dynamics of localized modes, their power $P$ plays the role of the effective mass \cite{kivshar2003optical,malomed_soliton_2022}.

In optics, the same (2+1)D NLSE as in Eq. (\ref{eq:refname4}) may also be posed as the model for the spatiotemporal propagation of light along axis $z$ in a planar waveguide with transverse coordinate $x$, while $y$ plays the role of the temporal coordinate, provided that the material dispersion is anomalous \cite{kivshar2003optical,malomed_soliton_2022}. In that case, the vortical structure carried by the necklace mode represents the \textit{spatiotemporal vorticity} \cite{dror_symmetric_2011}, which has recently drawn much interest in its general form \cite{chelpanova_spatiotemporal_2019,bliokh_spatiotemporal_2021,porras_transverse_2023}.

The same equation (\ref{eq:refname4}), with $z$ replaced by time $t$, represents the scaled form of the Gross-Pitaevskii equation (GPE) for the mean-field wave function $\psi $ of quasi-two-dimensional Bose-Einstein condensates with the attractive intrinsic nonlinearity \cite{malomed_soliton_2022}. Thus, the field configuration addressed here can be realized in various physical settings. Furthermore, the GPE can be introduced for the quasi-2D condensate created on a curved surface, rather than the flat plane \cite{craps_maximally_2017,tononi_bose-einstein_2019}. For instance, in the simplest case of a conical surface with angle $\alpha $ between its axis and the surface, the GPE similar to the one in Eq. (\ref{eq:refname4}), if written in the projection onto the horizontal plane (the one which is perpendicular to the conical axis) with polar coordinates $\left( r,\theta \right) $, takes the form of
\begin{equation}
i\frac{\partial \psi }{\partial {t}}+\frac{1}{2}\left( \frac{1}{r^{2}}\frac{%
\partial ^{2}\psi }{\partial \theta ^{2}}+\sin ^{2}\alpha \cdot \frac{%
\partial ^{2}\psi }{\partial \rho ^{2}}\right) +|\psi |^{2}\psi =0.
\label{conus}
\end{equation}%
Obviously, Eq. (\ref{conus}) carries over into the usual GPE in the limit of $\alpha =0$.

The beam's evolution is characterized by the Rayleigh (diffraction) length. In physical units, it is $l_{\mathrm{R}}=2\pi nR_{0}^{2}/\lambda $ \cite{zhang2016propagation}, where $n$ is the linear refractive index, $\lambda $ and $R_{0}$ being the wavelength and scaled transverse width, respectively. If the nonlinear term is ignored in Eq. (\ref{eq:refname4}), then the Gaussian beam $\psi (z=0,x,y)=\exp [-(x^{2}+y^{2})/2]$ expands by $\sqrt{2}$ passing $z=l_{\mathrm{R}}$ \cite{soljacic_integer_2001}.

In terms of the polar coordinates, the stationary light-beam solution of the NLSE can be written as $\psi (z=0,x,y)=f(r)\exp (iM\theta )$, where $M$ is the integer (positive or negative) topological charge (winding number). For this ansatz, Eq. (\ref{L2}) yields the universal relation between the power and angular momentum for stationary states:
\begin{equation}
L=MP,  \label{L=MP}
\end{equation}
the angular momentum per photon being $M\hbar $ in physical units \cite{franke-arnold_advances_2008}.

It is instructive to compare $L/P$ as an integer eigenvalue in quantum mechanics and classical optics. In the former case, the solutions of the linear Schr\"{o}dinger equation are characterized by the quantized angular momentum, which is an integer number times $\hbar $, while integral $\iint |\psi |^{2}dxdy$ represents the total probability, which is normalized to be $1$ (unlike the power in optics). The quantization of the angular momentum is similar to the fact that, in terms of optics, $L/P$ is the
above-mentioned integer winding number for stationary solitons. Nevertheless, the \textit{reduced angular momentum} $L/P$ borne by a necklace is not necessary an integer, unlike the case of stationary axisymmetric solitons.

Thus, we aim to consider the propagation dynamics for self-trapped structures carrying, generally speaking, noninteger values of $L/P$, by means of the PINN method. Therefore, we first define the residual neural network $f(z,x,y)$ as
\begin{equation}
f(z,x,y)\equiv i\frac{\partial \widehat{\psi }}{\partial {z}}+0.5\left( \frac{\partial ^{2}\widehat{\psi }}{\partial x^{2}}+\frac{\partial
^{2}\widehat{\psi }}{\partial y^{2}}\right) +|\widehat{\psi }|^{2}\widehat{%
\psi },  \label{eq:refname5}
\end{equation}
where $\widehat{\psi }(z,x,y)=\widehat{u}(z,x,y)+i\widehat{v}(z,x,y)$ is the predicted solution, cf. the first line in Eq. (\ref{eq:refname4}). Accordingly, $f(z,x,y)$ splits into the real and imaginary components:
\begin{equation}
f_{\mathrm{re}}=-\widehat{v}_{z}+0.5\nabla^2 \widehat{u}+(\widehat{u}%
^{2}+\widehat{v}^{2})\widehat{u}, \;
f_{\mathrm{im}}=-\widehat{u}_{z}+0.5\nabla^2 \widehat{v}+(\widehat{u}%
^{2}+\widehat{v}^{2})\widehat{v}.
\label{eq:refname6}
\end{equation}%

The predicted solution is added to PINN through Eq.~(\ref{eq:refname6}), which avoids overfitting by physical constraints, thus providing the appropriate physical interpretation for $\widehat{\psi }(z,x,y)$. The shared parameters of the neural networks $\widehat{\psi }(z,x,y)$ and $f(z,x,y)$ can be found by minimizing the MSE loss:
\begin{equation}
\mathcal{L}=\mathcal{L}_{\mathrm{IC}}+\mathcal{L}_{\mathrm{BC}}+\mathcal{L}_{%
\mathrm{f}},  \label{eq:refname7}
\end{equation}
where $\mathcal{L}_{\mathrm{IC}}$ is built from the computed difference between the incident data $\psi (z=0,x_{0}^{j},y_{0}^{j})$ and the predictable results $\widehat{\psi }(z=0,x_{0}^{j},y_{0}^{j})$ in the framework of PINN, $\mathcal{L}_{\mathrm{BC}}$ accounts for the periodic BC, and randomly chosen points in the entire spatial domain contribute to the $\mathcal{L}_{\mathrm{f}}$. All the loss terms can be realized, defining the MSEs as
\begin{equation}
\begin{split}
& \mathcal{L}_{\mathrm{IC}}=\frac{1}{n_{\mathrm{IC}}}\sum_{j=1}^{n_{\mathrm{%
IC}}}|\widehat{\psi }(x_{0}^{j},y_{0}^{j},0)-\psi
(x_{0}^{j},y_{0}^{j},0)|^{2}, \\
& \mathcal{L}_{\mathrm{BC}}=\frac{1}{n_{\mathrm{BC}}}\sum_{j=1}^{n_{\mathrm{%
BC}}}[|\widehat{\psi }(x_{1},y_{BC}^{j},z_{BC}^{j})-\psi
(x_{2},y_{BC}^{j},t_{BC}^{j})|^{2} \\
& \qquad +|\widehat{\psi }(x_{BC}^{j},y_{1},t_{BC}^{j})-\psi
(x_{BC}^{j},y_{2},z_{BC}^{j})|^{2}], \\
& \mathcal{L}_{\mathrm{f}}=\frac{1}{n_{f}}%
\sum_{j=1}^{n_{f}}|F(x_{f}^{j},y_{f}^{j},z_{f}^{j})|^{2},
\end{split}
\label{eq:refname8}
\end{equation}%
where the observable measurements $\{{\widehat{\psi }(0,x_{0}^{j},y_{0}^{j})}\}_{j=1}^{n_{\mathrm{IC}}}$ denote the sampled data utilized in the PINN at $z=0$, as provided by the IC. Further, $n_{\mathrm{IC}}$ and $n_{\mathrm{BC}}$ denote the number of training points on the initial and boundary conditions, respectively, and $n_{f}$ is the number of collocation points. Here $\{\widehat{\psi }(z_{\mathrm{BC}}^{j},x,y_{\mathrm{BC}}^{j})\}_{j=1}^{n_{\mathrm{BC}}}$ and $\{\widehat{\psi }(z_{\mathrm{BC}}^{j},x_{\mathrm{BC}}^{j},y)\}_{j=1}^{n_{\mathrm{BC}}}$ stand for the latent necklace solution related to randomly selected boundary training data, while $\{z_{f}^{j},x_{f}^{j},y_{f}^{j}\}_{j=1}^{n_{f}}$ are associated with the chosen points in the spatial domain for training the residual $f(z,x,y)$. The pseudo-spectral method \cite{yang2010nonlinear} is used to generate the training data, and the sampling points for the neural networks are obtained by the space-filling Latin Hypercube Sampling strategy (LHS) \cite{stein_large_1987}. The objective of the present work is to minimize the specific loss function as much as possible, that should allow one to produce high-resolution predictable necklace beams $\widehat{\psi }(z,x,y)$ close to the exact ones ${\psi }(z,x,y)$.

%%%%%%%%%%%%%%%%%%%%%%%%%%%%%%%%%%%%%%%%%%%%%%%%%%%%%%%%%%%%%%%%

\section{Data-driven necklace solitons}\label{sec3}

In this section, we apply the PINN method to the propagation dynamics of the necklace-ring quasi-solitons carrying integer, half-integer, or reduced angular momentum $L/P$. The input beam is taken as
\begin{equation}
\psi (z=0,x,y)=f(r)\cos (N\theta )\exp (iM\theta )  \label{psi}
\end{equation}%
with integers $N$ and $M$. Factor $\exp (iM\theta )$, where, as mentioned above, integer $M$ may be positive or negative, implies the addition of the angular momentum (vorticity) to the necklace. The angular momentum of ansatz (\ref{psi}), given by Eq. (\ref{L2}), amounts to the same general relation (\ref{L=MP}) as stated above.

The necklace beam adopted in the form of ansatz (\ref{psi}) has an annular structure, with the overall radius which is essentially larger than the radial thickness (see, in particular, Eq. (\ref{input}) below). The intensity is periodically modulated in the azimuthal direction by factor $\cos (N\theta )$, which helps one to avoid the onset of angular modulational instability of axisymmetric patterns. This input may be considered as a superimposition of two rings with identical azimuthally uniform intensity shapes and opposite topological charges, resulting in the interference pattern in the azimuth direction. Generally, the necklace exhibits slow radial expansion in the course of the propagation. The expansion is a consequence of a net radial force exerted on each bright spot (\textquotedblleft bead\textquotedblright ) in the necklace. However, this expansion evolves much slower than the usual diffraction. As shown below (see Figs. \ref{fig4} and \ref{fig5}(c)), in some cases the necklace-shaped input develops very slow shrinkage, rather than expansion.

Thus, the light beam preserves the necklace-like shape in the course of the long propagation. In particular, there is no significant change in the diameter of the necklace's ring and size of individual beads in the ring. Indeed, it has been previously reported that the necklace may persist over the propagation distance exceeding $50l_{\mathrm{R}}$, provided that the topological charge $M$ is smaller than the modulation number $N$, see Eq. (\ref{psi}) \cite{soljacic_integer_2001}. The angular momentum exhibits itself by inducing rotation of the necklace in the course of its propagation. The apparent stability of the self-trapped necklaces is actually a unique feature, in comparison to other soliton complexes. For this reason, we choose the necklace self-trapped beams in the Kerr nonlinear media to verify the validity of the PINN, aiming, in particular, to address effects of the angular momentum and centrifugal force.

We now direct attention to the initial-boundary-value problem for the necklace beams, applying the pseudo-spectral numerical integration to generate a high-resolution training dataset for predicting the nonlinear dynamics of the necklace beams by PINN. To meet the constraints of the initial-boundary conditions and governing equations, $n_{0}$ data points are randomly sampled inside the initial condition, $\psi (z=0,x,y)$, and periodic boundary conditions, $n_{f}$ values of auxiliary coordinates in the spatial region being sampled via the Latin Hypercube Sampling strategy. In the framework of the PINN setup shown in Fig.~\ref{fig1}, spatial coordinates $(z,x,y)$ are used to calculate the loss terms in Eq.~(\ref{eq:refname8}). After multiple iterations, the propagation dynamics of the necklace beams can be predicted with high accuracy.

%%%%%%%%%%%%%%%%%%%%%%%%%%%%%%%%%%%%%%%%%%%%%%%%%%%%%%%%%%%%%%%%%%%%%%%%%%%%%%

\subsection{Necklace beams with integer reduced orbital angular momentum}

First, we consider the necklace beams carrying an integer reduced angular momentum $L/P=M$, with the respective input in the form of
\begin{equation}
\psi (z=0,x,y)=\mathrm{sech}(r-r_{0})\cos (N\theta )\exp (iM\theta )
\label{input}
\end{equation}%
(cf. expression (\ref{psi})), with $r_{0}$ large enough, to make the necklace sufficiently narrow in the radial direction. When the vorticity is absent ($M=0$), the azimuthal modulation factor $\cos \left( N\theta \right)$ in Eq. (\ref{input}) splits the ring $r=r_{0}$ in a set $2N$ \textquotedblleft beads\textquotedblright , with angular separation $\Delta \theta =\pi /N$ between adjacent ones and opposite signs in them, i.e., with phase shift $\Delta \varphi =\pi $ between the adjacent beads. The effective mass of each bead, i.e., its power, is
\begin{equation}
P_{\mathrm{bead}}=P/(2N).  \label{bead}
\end{equation}%
If $M\neq 0$ in Eq. (\ref{input}), the phase shift between the nearest beads in the necklace is $\Delta \varphi =\pi +2\pi M/N$. It determines interaction forces between the beads in the ring. Namely, for
\begin{equation}
\pi /2<\Delta \varphi \equiv \pi +2\pi M/N<3\pi /2(\bmod2\pi ),
\label{repulsion}
\end{equation}%
there is a net radial outwards-directed force acting onto each bead, due to its \emph{repulsive interactions} with the nearest neighbors \cite{kivshar2003optical,malomed_soliton_2022}. Otherwise, i.e., in the cases of
\begin{equation}
0\leq \Delta \varphi \equiv \pi +2\pi M/N<\pi /2~~\mathrm{or~~} \; 3\pi /2<\pi +2\pi M/N\leq 2\pi (\bmod2\pi ),
\label{attraction}
\end{equation}%
the interaction between the nearest neighbors is attractive. In the latter case, the interaction forces tend to drive shrinkage of the necklace; however, this trend is counteracted by the additional centrifugal force applied to each bead, as per Eq. (\ref{centri}), see below. Following Ref. \citenum{dong_necklace_2021}, in the following analysis we set $r_{0}=6.83$ in Eq. (\ref{input}).

Implementing the process described above, the spatial domain $(z,x,y)\in \lbrack 0,3l_{\mathrm{R}}]\times \lbrack -18,+18]\times \lbrack -18,+18]$ was employed to reveal the beam's evolution, with the longitudinal and transverse coordinate steps $\Delta z=0.005$ and $\Delta x=\Delta y=0.45$, respectively. We use the six-hidden-layer neural network with $80$ neurons per hidden layer to approximate the latent solution $\psi (z,x,y)$, and the nonlinear activation function $\sigma $, as mentioned above, is chosen as $\tanh$ to characterize all test scenarios. The output of the $k$-$\mathrm{th}$ layer in the neural network is given by: $\phi ^{k} =\tanh(W^k\phi ^{k-1}+b^k)$. The training set is composed of randomly sampled $n_{0}=2500$ points for the initial-boundary conditions, and the number of collocation points is $n_{f}=20000$ for the PINN $f(z,x,y)$ defined as per Eq.~(\ref{eq:refname5}). The number of iteration steps is $10000$ for both the Adam and several-step L-BFGS optimization methods. Note that the L-BFGS optimization algorithm makes the network converging faster with smaller fluctuations, thus accelerating the PINN learning. The L-BFGS optimization ends when the following convergence condition for the iterations is achieved:
\begin{equation}
\frac{|\mathrm{Loss}^{n}-\mathrm{Loss}^{n-1}|}{\max \{|\mathrm{Loss}^{n}|,|
\mathrm{Loss}^{n-1},1|\}}<1.0\times \mathrm{np.finfo(float).eps},
\label{eq:refname9}
\end{equation}
where $\text{Loss}\mathrm{^{n}}$ stand for the loss value of the $n$-$\mathrm{th}$ iteration for the L-BFGS algorithm, and $1.0\times \mathrm{np.finfo(float).eps}$ is the Machine Epsilon. All training sets are conducted on a desktop computer equipped with an Intel(R) Xeon(R) CPU E5-2690 v4 @2.60GHz, NVIDIA GeForce RTX 2080 Ti.

\begin{figure}[tbph]
\centering
\includegraphics[width=0.45\textwidth]{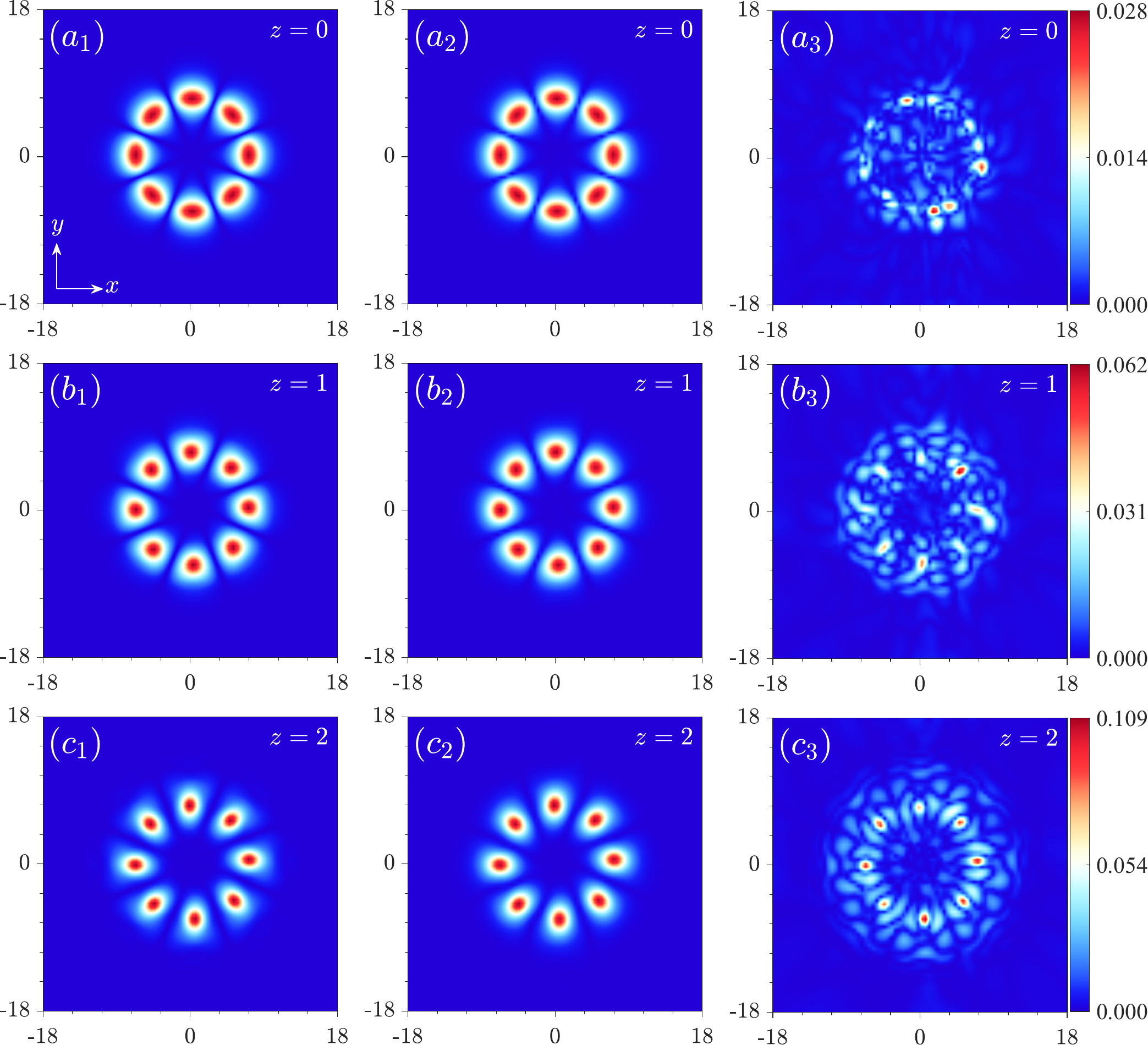}% Here is how to import EPS art
\caption{The evolution of the input necklace beam (\protect\ref{input}), taken as $\protect\psi (z=0,x,y)=\mathrm{sech}(r-6.83)\cos (4\protect\theta )\exp (i\protect\theta )$. ($a_{1}-c_{1}$): Results of the numerical simulations; ($a_{2}-c_{2}$): the data-driven solution produced by PINN at $z=0$. ($a_{3}-c_{3}$): The absolute error between the numerically found and data-driven necklace beams at $z=0$, its values being $\sim $ $10^{-3}$. }
\label{fig2}
\end{figure}

\begin{figure}[tbph]
    \centering
    \includegraphics[width=0.35\textwidth]{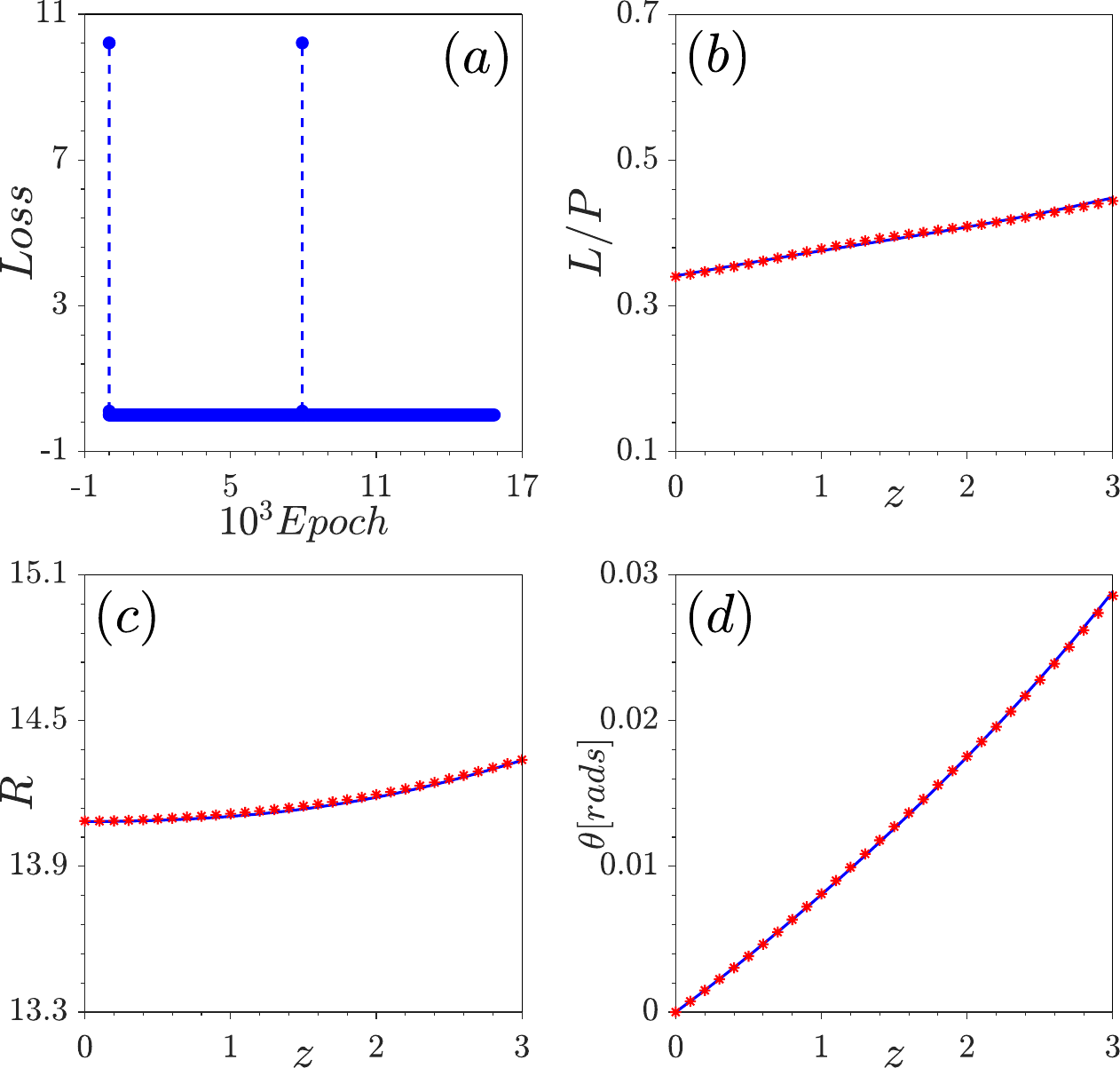}
    % Here is how to import EPS art
    \caption{(a) The loss function vs. the number of training epochs, for the same case as in Fig. \protect\ref{fig2}. (b) The dependence of $L/P$ on propagation distance $z$. (c,d) The effective radius $R$ [defined as per Eq. (\protect\ref{R})] and rotation angle of the necklace vs. the propagation distance. Blue lines and red asterisks represent, severally, results of the numerical simulations and their data-driven counterparts. }
    \label{fig3}
\end{figure}

Typical profiles of the necklace beams are displayed in Fig.~\ref{fig2}, for $N=4$ and $M=1$. A numerically exact necklace is obtained by means of the pseudo-spectral method of the numerical integration, as shown in Figs.~\ref{fig2}($a_{1}-c_{1}$). After $10000$ steps of the Adam iteration and several steps of L-BFGS training, the predicted solutions at $z=0,1,2$ are shown in Figs.~\ref{fig2}($a_{2}-c_{2}$). We find that the total loss value defined by Eq.~(\ref{eq:refname7}) is minimized to \textrm{2.2656692e-05}. The network has achieved the following values of the relative $\mathbb{L}_{2}$ error for $z=0,1,2$: \textrm{1.256216e-02, 3.149800e-02, 6.132568e-02}, respectively, in about $1478.6068$ seconds. The comparison at three different distances $z=0,1,2$ of the numerically exact necklace $\psi (z,x,y)$ and learning data $\widehat{\psi }(z,x,y)$ produced by the PINN is displayed in Figs.~\ref{fig2}($a_{3}-c_{3}$) by the absolute error, $|\widehat{\psi }(z,x,y)-\psi (z,x,y)|$. From the density map, one can conclude that results of the optimized PINN prediction matches the reference solution quite well in the entire spatial domain. The convergence curve of the loss function is presented in Fig.~\ref{fig3}(a), where the number of ``epochs'' represents the number of times that the whole body of the training data is fed into the neural network. It can be seen that the final convergence error of $\mathbb{L}_{2}$ is extremely small, indicating the excellent performance.

Although the annular beam gradually expands in the course of the propagation, it retains the stable necklace structure. The angular momentum is exhibited by the rotation of the entire necklace with angular velocity $\omega $, as seen in Fig.~\ref{fig3}(d), where the rotation angle of the expanding necklace is plotted as a function of the propagation distance (the rotation is not visible in Fig. \ref{fig2}, as $\omega $ is too small). The angular velocity of a narrow circular structure with radius $R$ can be estimated by equating the respective expression for the angular momentum, $PR^{2}\omega $ (recall power $P$ plays the role of the dynamical mass), to the above-mentioned expression $L=MP$ for the total angular momentum, i.e.,
\begin{equation}
\omega =M/R^{2}.  \label{omega}
\end{equation}%
This estimate approximately agrees with the numerical data presented in Fig.~\ref{fig3}(d).

The rotational motion of the bead implies the action of the centrifugal force, which must be taken into account when analyzing the dynamics of the necklace:
\begin{equation}
\mathcal{F}_{\mathrm{centrifugal}}=P_{\mathrm{bead}}\cdot R^{2}\omega \equiv
MP/\left( 2N\right) ,  \label{centri}
\end{equation}%
where the effective bead's mass (power) is substituted as per Eq. (\ref{bead}). Additional numerical results demonstrate that the necklaces with a larger winding number (vorticity) $M$ expand faster than ones with smaller $M$, which is a manifestation of the action of the centrifugal force, as per Eq. (\ref{centri}).

Values of the angular momentum $L$ and energy $P$ have been calculated for different propagation distances. In Fig.~\ref{fig3}(b), one can see that the reduced momentum, $L/P$, exhibits a slow increase in the course of the propagation, due to the gradual loss of the necklace's power through emission of linear waves from the propagating beam, while the rate of the radiation loss of the momentum is essentially lower.

The exact definition of the radius (width) of the necklace beam is adopted as
\begin{equation}
R=P^{-1}{\iint r|\psi |^{2}dxdy}.  \label{R}
\end{equation}%
Note that, in terms of the necklace's moment of inertia, which is $I\sim PR^{2}$, the angular momentum is $L=\omega I$. As seen in Fig. \ref{fig3}(c), the radius is slowly increasing in the course of the evolution, due to the weak expansion of the necklace under the action of the weak interaction of each \textquotedblleft bead\textquotedblright with its nearest neighbors in the necklace. We stress that the expansion is much slower than the usual diffraction-driven expansion, allowing the beam to maintain the necklace shape. In the case displayed in Figs. \ref{fig2} and \ref{fig3} the specific slowness of the expansion is explained by the fact that, for $N=4$ and $ M=\pm 1$, the phase shift $\Delta \varphi $ coincides with the boundary between the expansion and shrinkage intervals (\ref{repulsion}) and (\ref{attraction}). It may be expected that the necklace will seem as a practically stationary (non-expanding) pattern in the experiment. Figure \ref{fig3}(b) demonstrates that the effective stability of the necklace beam is accurately predicted by the PINN method.

\begin{figure}[tbph]
    \centering
    \includegraphics[width=0.45\textwidth]{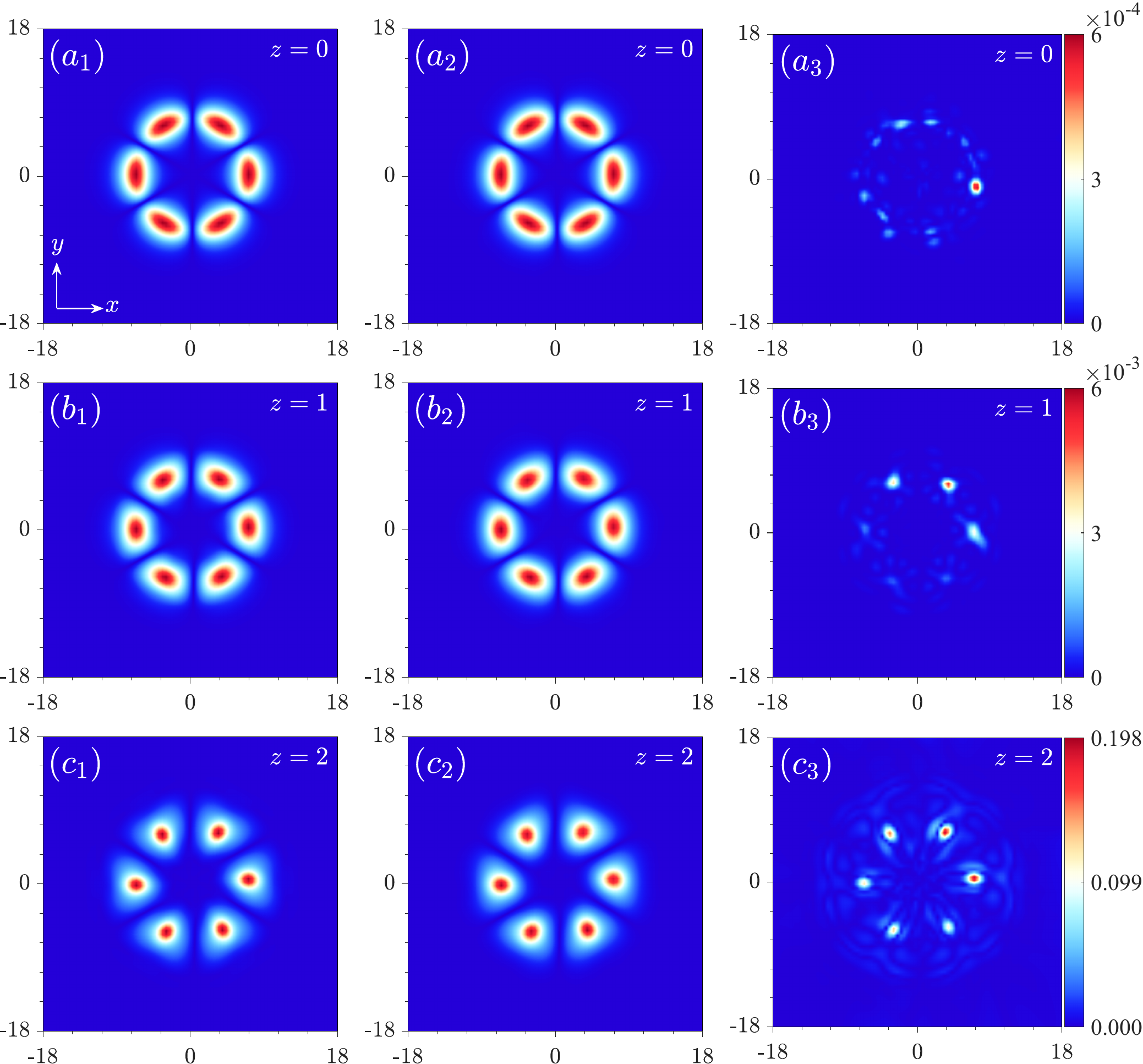}
    % Here is how to import EPS art
    \caption{(a) Results for the input beam $\protect\psi (z=0,x,y)=\mathrm{sech}(r-6.83)\cos (3\protect\theta )\exp (i\protect\theta )$. ($a_{1}-c_{1}$): The numerically found necklace beam; ($a_{2}-c_{2}$) the data-driven necklace solutions produced by the PINN at $z=0$. ($a_3-c_3$): The absolute error between the exact solution and predicted solutions.}
    \label{fig4}
\end{figure}

As mentioned above, the necklace's angular momentum, which is proportional to $\omega R^{2}$, remains practically conserved. Accordingly, the angular velocity $\omega $ of the rotation does not stay strictly constant, slowly decaying $\sim M/R^{2}$ [as per Eq. (\ref{omega})] with the very slow increase of the necklace's radius $R$.

In the case of the dominant attractive interaction between the neighboring beads, the necklace beams are displayed in Fig. \ref{fig4}, for $N=3$ and $M=1$. Note that this values yield $\Delta \varphi =5\pi /6$, which indeed corresponds to the attraction between neighboring beads, pursuant to Eq. (\ref{attraction}). The corresponding numerical necklace solution is obtained by the pseudo-spectral method, as shown in Figs. \ref{fig4}($a_{1}-c_{1}$). After $10000$ steps of Adam training and L-BFGS-B algorithm, the predicted solutions at $z=0,1,2$ are shown in Figs. \ref{fig4}($a_{2}-c_{2}$). The total loss value is minimized to $1.138714e-02$. The relative $\mathbb{L}_{2}$ errors are $9.912175e-03,3.224339e-02,6.263887e-02$, respectively, for $z=0,1,2$. The absolute error between numerically exact solutions $\psi (z,x,y)$ and predicted solutions $\widehat{\psi }(z,x,y)$ is displayed in Figs.~\ref{fig4}($a_{3}-c_{3}$).

\begin{figure}[tbph]
    \centering
    \includegraphics[width=0.35\textwidth]{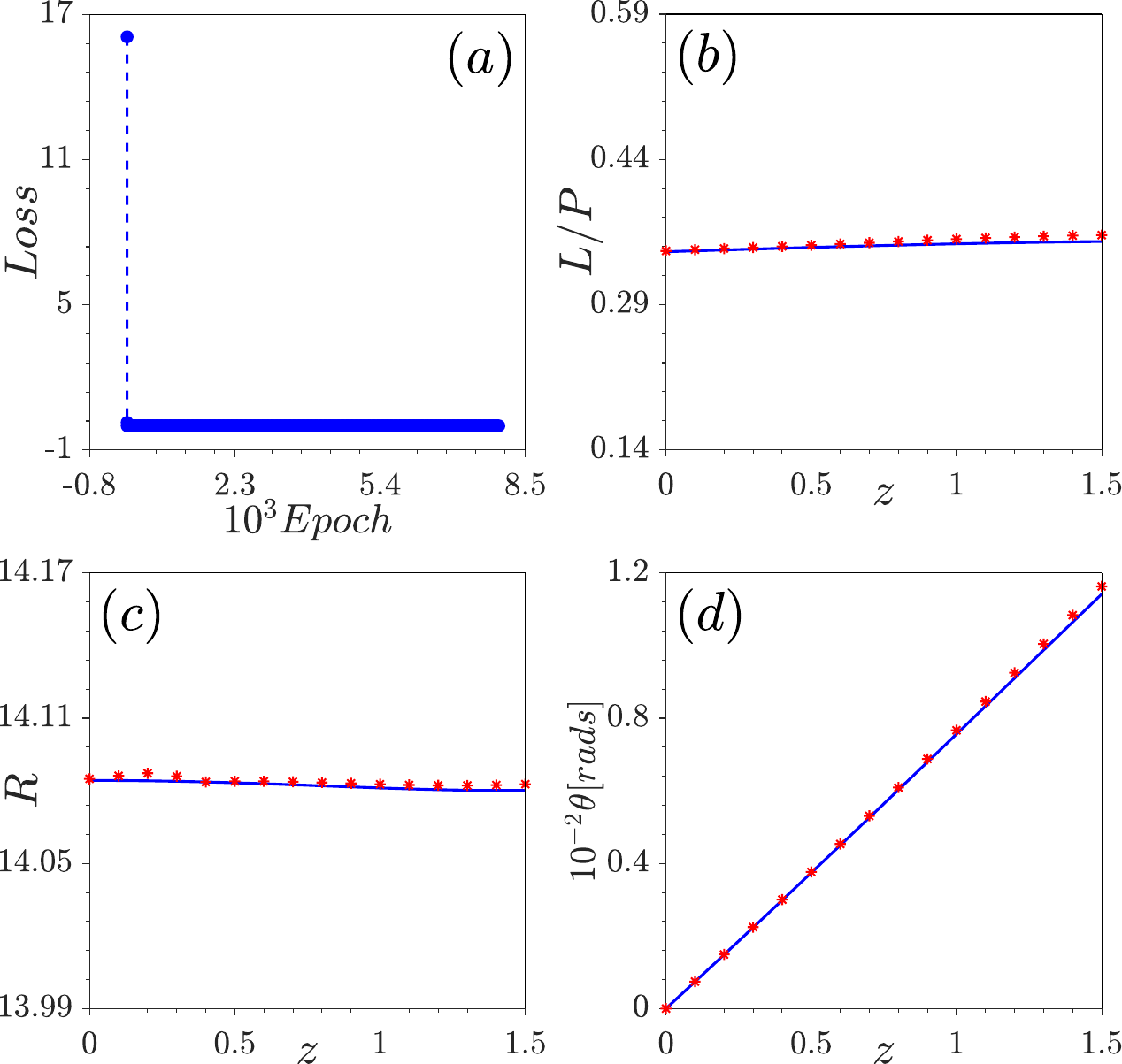}
    % Here is how to import EPS art
    \caption{(a) The loss function vs. the number of training epochs. (b) The dependence of $L/P$ on propagation distance $z$. (c,d) The effective radius $R$ and rotation angle $\protect\theta $ of the necklace vs. the propagation distance. Blue lines and red asterisks represent, severally, results of the numerical simulations and their data-driven counterparts.}
    \label{fig5}
\end{figure}

The convergence curve of the loss function, corresponding to the case presented in Fig. \ref{fig4}, is displayed in Fig.~\ref{fig5}. Unlike the necklace where neighboring beads repel each other, in this case the necklace's radius is slowly decreasing in the course of the evolution, as seen in Fig.~\ref{fig5}(c). The predicted results for $L/P$ and the rotation velocity are displayed in Figs.~\ref{fig5}(b) and (d). Thus, PINN is able to correctly predict the slow shrinkage of the necklace in the present case.

It is well known that the neural network structure is not always the same for different problems. In general, the PINN with more hidden layers and neurons per layer shows better performance. Yet, we have found that, when the number of neurons per layer exceeds $80$, there is little impact on the performance of the model after several tests. In addition, when the number of hidden layers (the PINN depth, $\mathrm{d}$) exceeds $6$, larger storage space and longer training time lead only to a small improvement in the model's performance. In Table \ref{table:1} we compare different items (PINN characteristics), including the training time, loss value and $\mathbb{L}_{2}$ error at different distances, to exhibit the improvement of each item with the increase of the depth. It is concluded that PINN with fewer layers can also produce relatively accurate solutions, which indicates the great advantage of the PINN method in predicting the beam evolution. Comparing networks with different numbers of hidden layers, we have found that PINN with $6$ hidden layers has the smallest loss value and $\mathbb{L}_{2}$ error, and the iteration time also shows advantages in this case. Considering the reasonable balance between the precision and the method's complexity, we chose a network with $6$ hidden layers to explore the evolution of the necklace with the integer reduced orbital angular momentum. Following this, the same number of the layers is selected below to predict the dynamics of the necklace beams with half-integer and fractional values of the reduced angular momenta. In the latter case, the learning results of $\widehat{\psi }(z,x,y)$ attains accuracy $\sim 10^{-2}$.

\begin{table*}[htbp]
\renewcommand\arraystretch{1}
% \scalebox{0.4}{}{
\caption{The comparison between results produced by PINN with different
depths ($\mathrm{d}$), i.e., numbers of hidden layers.}
\label{table:1}\centering
\fontsize{10}{14} \selectfont    %{字体尺寸}{行距}
\begin{tabular}{l|cccc}
\toprule \diagbox [width=10em,trim=l] {Data}{Depths(d)} & 1 & 2 & 4 & 6 \\
\hline
Training time & 296 s & 1889 s & 1608 s & 1559 s \\
Loss value & 3.33478e-2 & 3.514235e-4 & 2.292846e-5 & 2.265669e-5 \\
$\mathbb{L}_{2}$ error ($z=0$) & 9.552168e-1 & 5.808277e-2 & 1.255248e-2 &
1.256216e-2 \\
$\mathbb{L}_{2}$ error ($z=1$) & 9.559227e-1 & 1.216673e-1 & 3.303233e-2 &
3.149800e-2 \\
$\mathbb{L}_{2}$ error ($z=2$) & 9.591696e-1 & 2.624547e-1 & 6.386885e-2 &
6.132568e-2 \\
\bottomrule
\end{tabular}
\vspace{0cm}
    % }
\end{table*}

%%%%%%%%%%%%%%%%%%%%%%%%%%%%%%%%%%%%%%%%%%%%%%%%%%%%%%%%%%%%%%%%%%%%%%%%%%%%%%%%%%%%

\subsection{Necklace beams with half-integer values of the reduced angular momentum}

As mentioned above, for azimuthally non-uniform modes the reduced angular momentum, $L/P$, may take non-integer values. Here, following Refs. \citenum{soljacic_integer_2001,dong_necklace_2021}, we construct necklace beams with half-integer values of $L/P$, using, instead of input (\ref{input}), the one
\begin{equation}
\psi (z=0,x,y)=0.5f(r)\left[ \exp \left( i(M+N)\theta \right) +\exp
\left( -iN\theta \right) \right] ,  \label{half}
\end{equation}%
with integer numbers $M$ and $N$. Indeed, definition (\ref{L}) of the angular momentum, applied to this expression, yields $L/P=M/2$, each photon contributing $M\hbar /2$ to the expectation value of the total angular momentum. The azimuthal intensity pattern corresponding to Eq. (\ref{half}) is
\begin{equation}
\left\vert \psi (z=0,x,y)\right\vert ^{2}=f^{2}(r)\cos ^{2}\left[
(M/2+N)\theta \right] .  \label{cos}
\end{equation}%
It is obvious that this expression (\ref{cos}) with an odd integer value of topological charge $M$ represents a necklace composed of the odd number of the beads,
\begin{equation}
n_{\mathrm{bead}}=M+2N.  \label{nbead}
\end{equation}%
Typical profiles of the necklace with the half-integer angular momentum are displayed in Fig.~\ref{fig4}, for $M=1$, $N=3$, and $n_{\mathrm{bead}}=7$. As well as in the case of even $n_{\mathrm{bead}}$, the total power of these necklaces is evenly distributed between the beads.

\begin{figure}[tbph]
    \centering
    \includegraphics[width=0.45\textwidth]{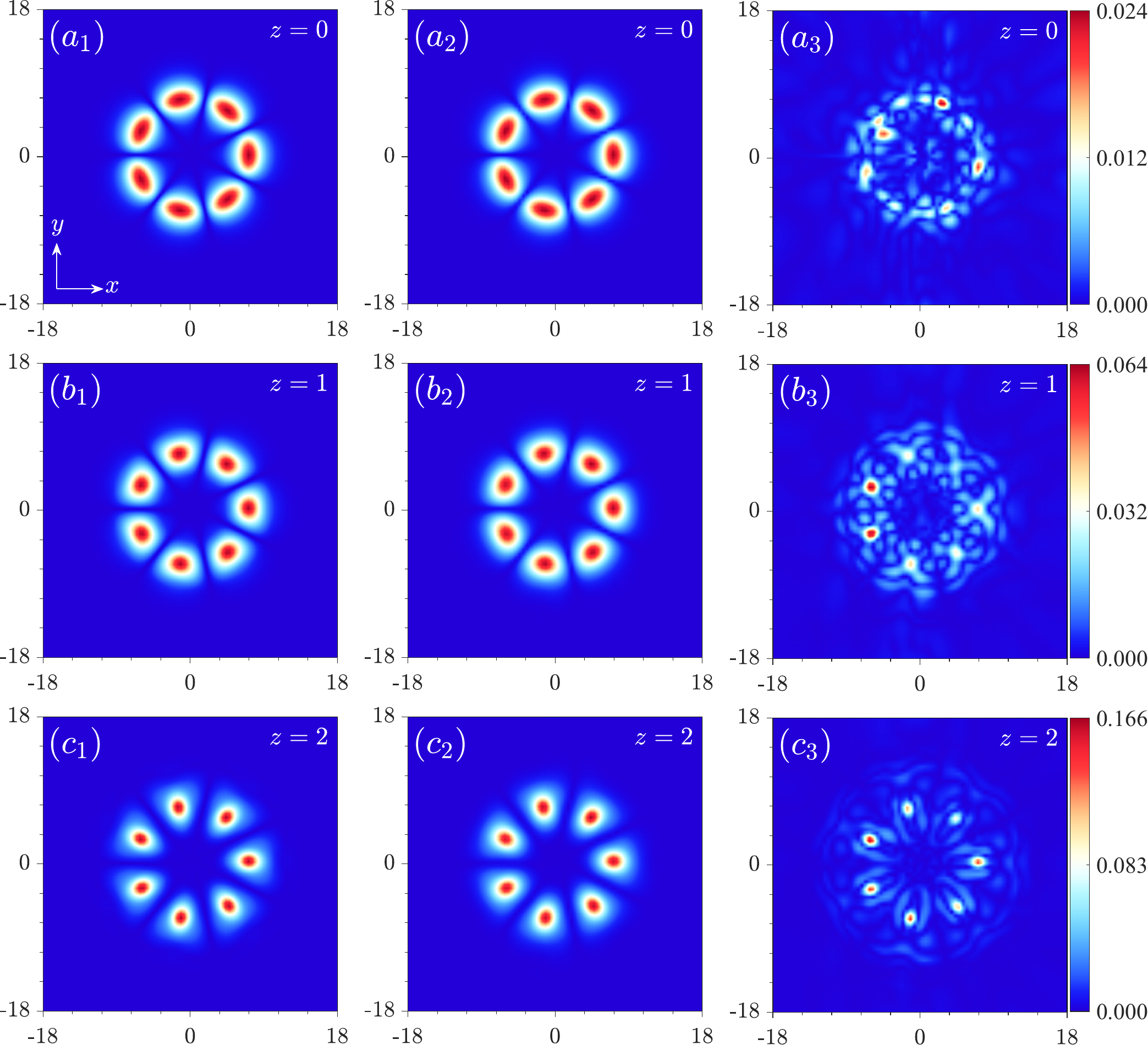}% Here is how to import EPS art
    \caption{The evolution of the input beam (\protect\ref{half}) representing the half-integer reduced angular momentum, with $M=1$, $N=3$, and the number of beads $n_{\mathrm{bead}}=7$, as per Eq. (\protect\ref{nbead} ), \textit{viz}., $\protect\psi (z=0,x,y)=\mathrm{sech}(r-6.83)[\exp (4i\protect\theta )+\exp (-3i\protect\theta )]/2$. ($a_{1}-c_{1}$) The necklace solitons produced by the numerical simulation. ($a_{2}-c_{2}$) The data-driven necklace solutions produced by the PINN. ($a_3-c_3$): The absolute error between the numerical and data-driven necklace beams. }
    \label{fig6}
\end{figure}

\begin{figure}[tbph]
    \centering
    \includegraphics[width=0.35\textwidth]{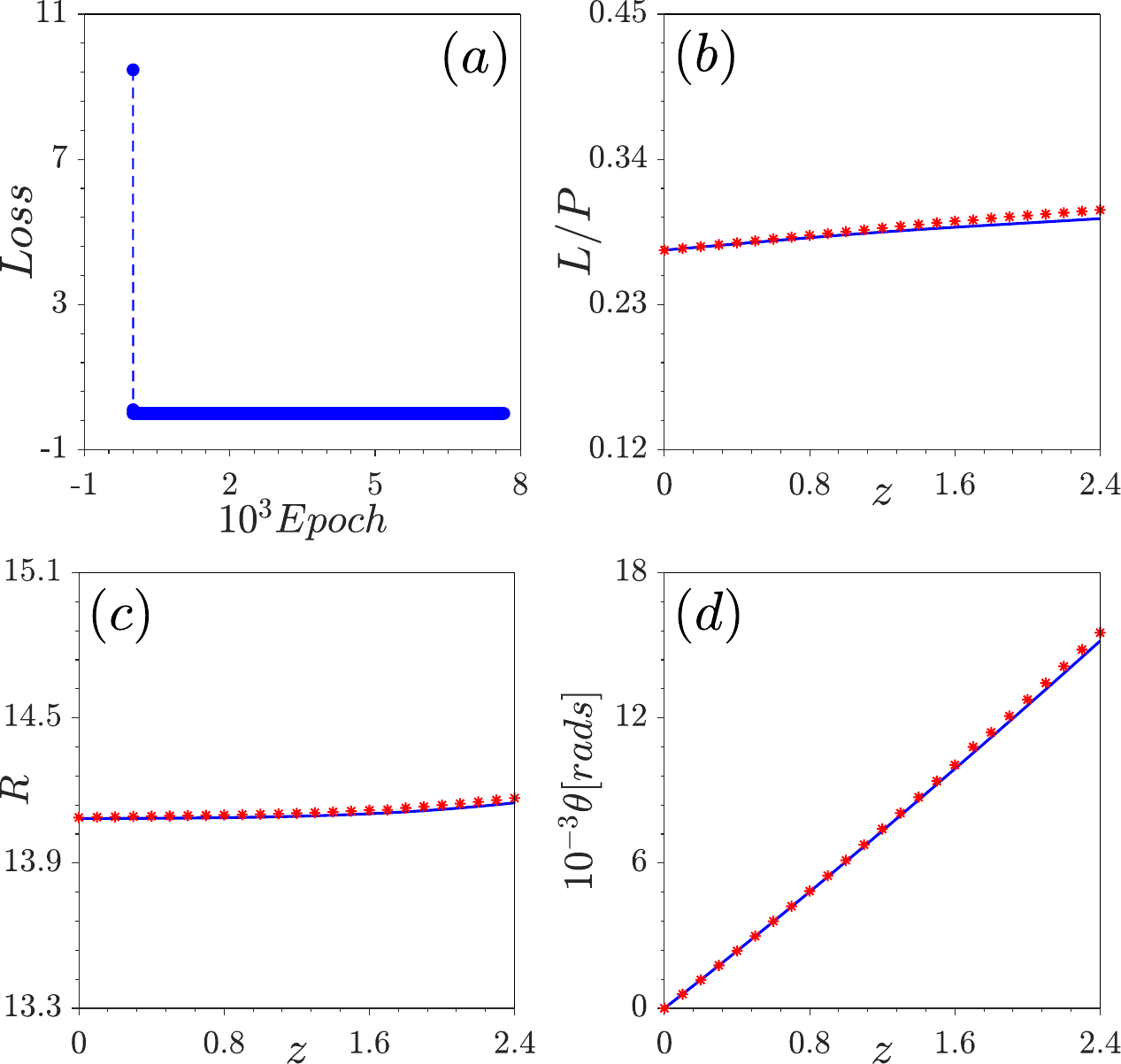}
    % Here is how to import EPS art
    \caption{Characteristics of the necklace beam with $L/P=1/2$. (a) The loss function for training vs. the number of epochs. (b) The dependence of $L/P$ on the propagation distance. (c,d) The effective radius $R$ and rotation angle of the necklace as a function of $z$. The blue lines and chains of red asterisks represent, respectively, the results produced by the numerical simulations and PINN method.}
    \label{fig7}
\end{figure}

As seen in Fig. \ref{fig6}, the evolution of the necklace beams with odd values of $n_{\mathrm{bead}}$, driven by the PINN method, is similar to that in case of integer $M$ and even $n_{\mathrm{bead}}$. In this case, we used the six-hidden-layer neural network with $80$ neurons per hidden layer, combined with a hyperbolic-tangent activation function \cite{sharma_activation_2020}, to characterize the latent solution $\psi (z,x,y)$ by minimizing the loss value. For this purpose, high-resolution training data was obtained by dint of the pseudo-spectral method, as shown in Fig.~\ref{fig6}($a_{1}-c_{1}$). In the entire spatial domain $(z,x,y)\in \lbrack 0,3l_{\mathrm{R}}]\times \lbrack -18,18]\times \lbrack -18,18]$, the
longitudinal coordinate step was $\Delta z=0.005$, and the transverse coordinates domain was divided into $80$ equal intervals in both the $x$ and $y$ directions. For the initial-boundary conditions, $39\%$ of the dataset is selected as sampling points. It is different from the previous cases in that $30000$ collocation sampled points are produced by means of the Latin Hypercube Sampling strategy in the whole spatial domains. After $10000$ steps of the Adam learning and L-BFGS optimization, we obtain the predicted necklace beams as shown in Fig.~\ref{fig6}($a_{2}-c_{2}$). The corresponding convergence process is plotted in Fig.~\ref{fig7}(a). The network achieves the following values of the relative error $\mathbb{L}_{2}$ for different distances, $z=0,1,2$, in about $1824.8113$ seconds: \textrm{1.169802e-02}, \textrm{3.147477e-02}, \textrm{6.167354e-02}, respectively, the total loss value being \textrm{2.1343205e-05}. In Fig.~\ref{fig6}($a_{3}-c_{3}$), we compare the training solution with the numerically found one at $z=0,1,2$. The corresponding absolute errors, $|\widehat{\psi }(z,x,y)-\psi (z,x,y)|$, are \textrm{2.400e-02, 6.400e-02, 1.660e-1},
respectively. The error gradually increases, resulting from the slow expansion of the propagating self-trapped necklace beams. Thus the PINN-produced predictions are sufficiently accurate as well for the
necklaces with the half-integer values of the reduced angular momentum.

To further highlight the prediction ability of the PINN in the case of half-integer initial values of $L/P$, we monitored the evolution of $L/P$ and phase of the necklace beam. As seen in Fig.~\ref{fig7}(b), the
expectation value of $L/P$ remains approximately constant, similar to the case of the integer initial value of $L/P$, cf. Fig.~\ref{fig3}(b).

The substitution of ansatz (\ref{half}) in expression (\ref{L2}) for the angular momentum yields the relation $L=MP/2$, with the extra factor of $1/2$, in comparison to the relation (\ref{L=MP}), which holds for ansatz (\ref{input}). Accordingly, the angular velocity of the rotating necklace is half of that given above by Eq. (\ref{omega}), \textit{viz}., $\omega =M/\left(2R^{2}\right) $, which is indeed relatively close to $\omega $ corresponding to Fig. \ref{fig7}(d). Further, the respective centrifugal force is smaller
by factor $1/4$, in comparison with the expression given by Eq. (\ref{centri}). Eventually, the weaker centrifugal makes the expansion of the necklace with the half-integer of value $L/P$ force extremely slow, as observed in Fig.~\ref{fig7}(c).

%%%%%%%%%%%%%%%%%%%%%%%%%%%%%%%%%%%%%%%%%%%%%%%%%%%%%%%%%%%%%%%%%%%%%%%%%%%%%%%%%%%%%%%%%

\subsection{Necklace beams with fractional values of the reduced angular momentum}

Here, we construct the necklace beam with an arbitrary fractional value of $L/P$, adopting the input
\begin{equation}
\psi (z=0,x,y) =f(r)[A\exp (iM\theta )+B\exp (iN\theta ) \quad +C\exp (-iQ\theta )],
\label{input2}
\end{equation}%
with integer numbers $M$, $N$, and $Q$, and
\begin{equation}
L/P=(A^{2}M+B^{2}N-C^{2}Q)/(A^{2}+B^{2}+C^{2}),  \label{fract}
\end{equation}%
as given by the substitution of ansatz (\ref{input2}) in Eq. (\ref{L2}). The necklaces constructed by means of this ansatz are stable if the parameters are set appropriately. In what follows below, we set $A=1$, $B=7$, $C=8$, $N=Q=8$, and $M=15$, the corresponding ratio given by Eq. (\ref{fract}) being $L/P=-35/38$, which indicates that each photon contributes $0.9211\hbar $ to the expected value of the total angular momentum.

\begin{figure}[tbph]
\centering
\includegraphics[width=0.45\textwidth]{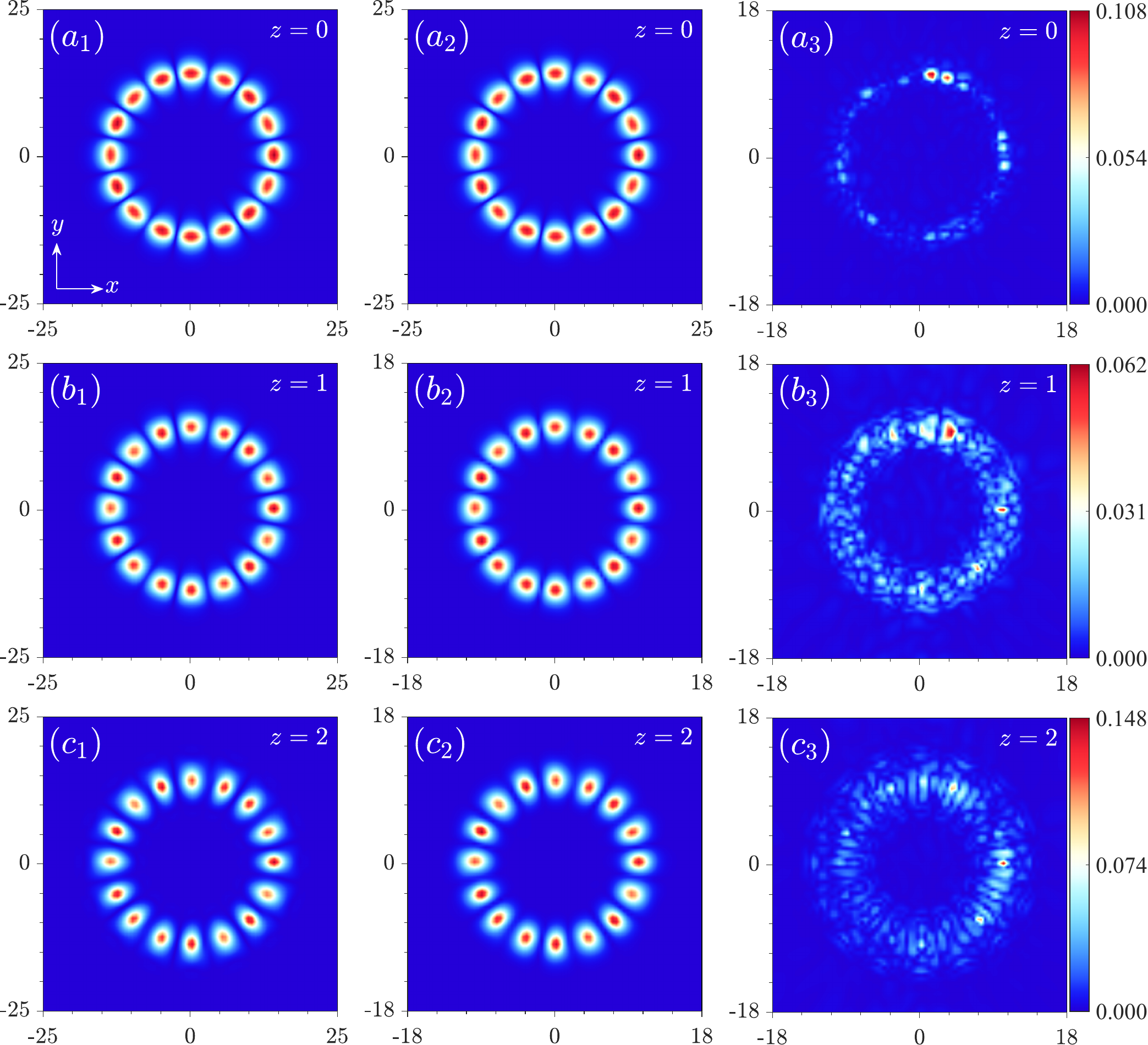}% Here is how to import EPS art
\caption{The input beam with the fractional angular momentum, taken, as per Eq. (\protect\ref{input2}), \textit{viz}., $\protect\psi (z=0,x,y)=\mathrm{sech}(r-13.66)[\exp (15i\protect\theta )+7\exp (8i\protect\theta )+8\exp (-8i\protect\theta )]/16$. ($a_{1}-c_{1}$) The evolution of the necklace
produced by the numerical simulations. ($a_{2}-c_{2}$) The evolution of the necklace produced by the PINN. ($a_3-c_3$): The absolute error between the numerical and data-driven necklaces. }
\label{fig8}
\end{figure}

The evolution of this necklace beams is shown in Fig.~\ref{fig8}. Although the shape of the necklace resembles those displayed above in Figs. \ref{fig2} and \ref{fig6}, it is different in that it exhibits the evolution of small azimuthal perturbations. This results in slight modulation of the multi-bead structure, whose intensity distribution is no longer uniform, and, accordingly, weak exchange in the power and angular momentum between adjacent beads is observed in the numerical simulations displayed in Figs.~\ref{fig8}($a_{1}-c_{1}$).

\begin{figure}[tbph]
\centering
\includegraphics[width=0.35\textwidth]{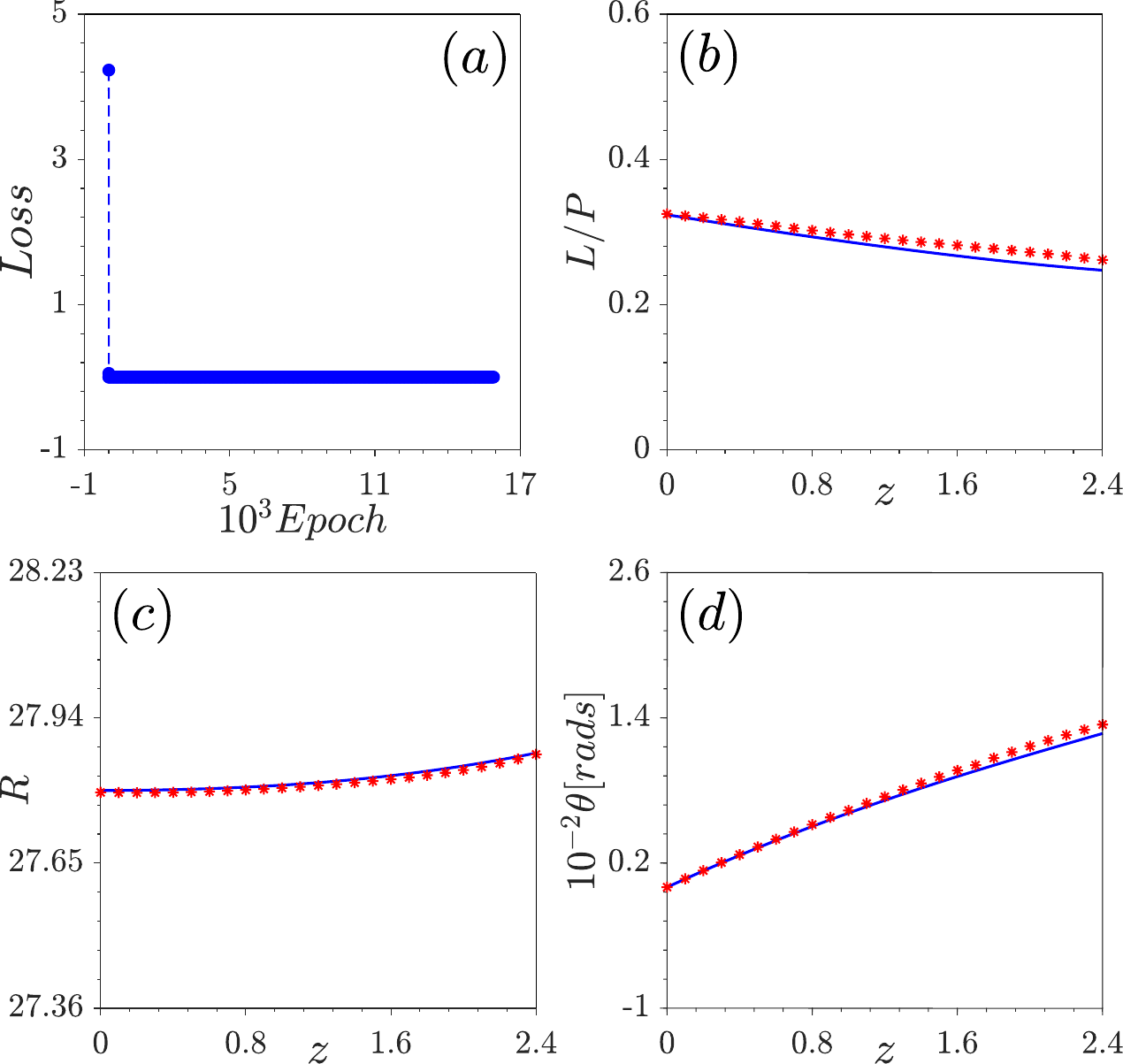}
% Here is how to import EPS art
\caption{The necklace beam with the fractional reduced angular momentum, $L/P=-35/38$. (a) The loss function of training vs. the number of epochs. (b) The dependence of $L/P$ on the propagation distance. (c,d) The effective radius $R$ and rotation angle of the necklace as a function of $z$. Blue lines and chains of red asterisks show the simulated and data-driven results, respectively. }
\label{fig9}
\end{figure}

To explore the accuracy of the PINN method in the present case, the set of sampling points is obtained by means of the pseudo-spectral method in the whole spatial domain, $(z,x,y)\in \lbrack 0,2]\times \lbrack -25,25]\times \lbrack -25,25]$. The transverse and longitudinal coordinate steps are taken as $\Delta x=\Delta y=0.45$ and $\Delta z=0.005$, respectively. The loss function given by Eq. (\ref{eq:refname7}) is learned via using the PINN framework consisting of the six-layer deep neural network with 60 neurons per layer. We randomly chose $2500$ points from the initial-boundary conditions, and $30000$ collocation points were deduced by the Latin Hypercube Sampling in the spatial domain. After $10000$ Adam iterations and the L-BFGS optimization, the network achieved the following values of the relative error $\mathbb{L}_{2}$ at $z=0,1,2$: \textrm{2.931236e-02, 3.559150e-02, 5.880743e-02}, respectively, in about $2185.5764$ seconds. The total loss is close to $2.4911773e-05$, the entire iteration process being documented in Fig.~\ref{fig9}(a).

The shapes of the necklace produced by PINN in this case are presented in Figs.~\ref{fig8}($a_{2}-c_{2}$). Comparing with the results of the numerical simulations, PINN can accurately predict the complex pattern and evolution dynamics of the necklace with the fractional reduced angular momentum. In the reported results, one can see that the symmetry of the multi-bead structure is indeed weakly broken, as mentioned above. In the course of the propagation, the necklace keeps its shape, with only slight relative changes between adjacent beads, which is due to the energy exchange between them (as also mentioned above). The comparison of the numerically exact necklace and training solution produced by the PINN method is
displayed in Figs.~\ref{fig8}($a_{3}-c_{3}$). Although the absolute error is relatively large at $z=0$, the prediction error somewhat improves in the course of the evolution. The error stays within reasonable margins, making the predicted results reliable.

If a necklace can stably propagate, the presence of the angular momentum is exhibited by the rotation of the necklace as a whole. For the necklace with the fractional values of the reduced angular momentum, the symmetry of the multi-bead structure changes, as mentioned above. Compared to the necklace beams with the integer or half-integer values, the present one, carrying the fractional reduced momentum, features the smallest rotation velocity in Fig.~\ref{fig9}(d). In the course of the propagation, although power $P$ and
angular momentum $L$ slowly decrease, the necklace remains stable. $P$ and $L$ are still approximately conserved because, compared to the initial values, shares of $P$ and $L$ taken away by the emission of radiation are negligible. Furthermore, PINN accurately predicts the radiation of $P$ and $L$ as the necklace propagates, as shown in Fig.~\ref{fig9}(b). Then, by calculating and predicting the effective radius of the beam, the retardation of the expansion of the present necklace beam is revealed, i.e., the effective radius barely changes in Fig.~\ref{fig9}(c). We can thus accurately predict the angular momentum and energy of the beam, as well as the entire propagation dynamics. It is seen that all the (blue) curves obtained from the numerical solutions are in good agreement with the chains of red asterisks generated by the PINN method.

%%%%%%%%%%%%%%%%%%%%%%%%%%%%%%%%%%%%%%%%%%%%%%%%%%%%%%%%%%%%%%%%%%%%%%%%%%%%%%%%%%%%

\section{Data-driven parameter discovery}\label{sec4}
Here we shift our attention to the consideration of the data-driven parameters discovery of the (2+1)D NLSE via the PINN scheme. This process is of great importance in identifying models for various applications. Discovering model parameters by means of traditional numerical methods is a challenging problem. Therefore, a promising approach is to infer unknown coefficients of physical significance from auxiliary measurements by leveraging the underlying physics. We here perform this for coefficients of the underlying equation in (\ref{eq:refname4}). Thus, the NLSE with undiscovered parameters is written as
\begin{equation}
    {i}\psi_{z} +\delta _{1}\left(\psi_{xx}+ \psi_{yy}\right)
    +\delta _{2}|\psi |^{2}\psi =0,  \label{eq:refname10}
\end{equation}%
where, as said above, $\psi (z,x,y)=u(z,x,y)+iv(z,x,y)$ is the normalized complex field amplitude of the beam. Parameters $\delta _{1}$ and $\delta _{2}$ stand for the diffraction and {Kerr} nonlinear coefficients, respectively. They are unknown real numbers to be found by means the PINN method.

Similar to the forward problem, we construct the neural network $\widehat{\psi }(z,x,y,\delta _{1},\delta _{2})$ with weights conforming to the normal distribution, to characterize complex field $\psi (z,x,y)$. The corresponding residual neural network is defined as
\begin{equation}
    F(z,x,y)\equiv i\psi_{z}+\delta _{1}\left(\psi_{xx}+ \psi_{yy}\right) +\delta _{2}|\psi |^{2}\psi .  \label{eq:refname11}
\end{equation}
It creates a bridge between the physical model and the residual neural network that contains intrinsic information, taking into regard that Eq.~(\ref{eq:refname10}) and the residual neural network defined by Eq.~(\ref{eq:refname11}) share the same parameters. Thus, we introduce the complex fields into the residual neural network, and the corresponding complex-values PINN $F(z,x,y)=F_{u}(z,x,y)+iF_{v}(z,x,y)$ is written as
\begin{equation}
 F_{u}=-v_z+\delta_{1}\nabla ^2 u+\delta_{2} (u^2+v^2)u, \; F_{v}=u_z+\delta_{1}\nabla ^2 v+\delta_{2}(u^2+v^2)v.
    \label{eq:refname12}
\end{equation}
To train the PINN $\widehat{\psi }(z,x,y,\delta _{1},\delta _{2})$ and obtain parameters ($\delta _{1},\delta _{2}$), the loss function is defined as
\begin{equation}
    \begin{split}
    \mathcal{L} &= \frac{1}{N} \sum_{j=1}^{N} [|\widehat{u} (z^{j},x^{j}, y^{j})
    - u^{j}|^2 + |\widehat{v} (z^{j},x^{j}, y^{j}) - v^{j}|^2 \\
    & \quad + |F_{u}(z^{j},x^{j}, y^{j})|^2 + |F_{v}(z^{j},x^{j}, y^{j})|^2],
    \end{split}
    \label{eq:refname13}
\end{equation}
where $\{z^{j},x^{j},y^{j},\psi ^{j}\}_{j=1}^{N}$ stands for randomly chosen samples obtained by means of the pseudo-spectral method. Note that the network structure of the inverse problem is similar to that of the forward problem, except that the loss function is different.

According to the previously considered forward problem, we use the necklace beam with the integer angular momentum as the dataset. The exact necklace solution for the (2+1)D equation~(\ref{eq:refname10}) with parameters $\delta _{1}=0.5$ and $\delta _{1}=1$ is employed to build sampling point sets. As the necklace beams maintain the quasi-stable propagation with slow expansion, the coefficients of the physical model are learned sequentially during the data-driven parameter discovery. First, we look for the diffraction coefficient $\delta _{1}$. Specifically, the PINN structure is designed as the six-layers deep neural network with $10$ neurons per layer, and the parameters are initialized as $\delta _{1}=0,\delta _{2}=1$. We select a $9.77\%$ dataset across the entire spatial regions $(z,x,y)\in \lbrack 0,3l_{\mathrm{R}}]\times \lbrack -18,18]\times \lbrack -18,18]$. Furthermore, the Adam optimizer is set with $20000$ steps and the L-BFGS optimization algorithm keeps $50000$ steps. Next, we analyze the Kerr nonlinear coefficient $\delta _{2}$ of Eq. (\ref{eq:refname10}). Unlike the training parameter $\delta _{1}$, only $1.56\%$ of the full dataset is selected as the training data. Another difference is the number of Adam steps, which is set to be $5000$. The same scenarios are used for training in the presence of noise.

The training results for the unknown parameters $\delta _{1},\delta _{2}$ are presented in Table \ref{tab:2}, including different situations and the training error. When noise is not added, the unknown parameters $\delta _{1},\delta _{2}$ are learned to be $0.49565$ and $0.98456$, while, in the presence of the $1\%$ noise, coefficients $\delta _{1},\delta _{2}$ are obtained as $0.50601$ and $0.97770$, respectively. Eventually, the relative errors of $\delta _{1}$ are $0.8\%$ and $1.2\%$, and those of $\delta _{2}$ are $1.5\%$ and $2.2\%$, respectively. Although the sampling set is perturbed by the noise, the unknown parameters can be correctly identified with high precision. Thus, we accurately predict the physical parameters of the model with fewer sampling points. This result is a relevant example of the enhancement offered by loading physical information into the neural network. It highlights the potential of this technique for solving high-dimensional inverse problems.
\begin{table*}[htbp]
    \renewcommand\arraystretch{1}
    \caption{Data-driven parameter discovery for parameters $\protect\delta _{1}$
    and $\protect\delta _{2}$ of Eq. (\protect\ref{eq:refname11}), and the
    corresponding errors.}
    \label{tab:2}\centering
    \fontsize{10}{14} \selectfont    %{字体尺寸}{行距}
    \begin{tabular}{l|cccc}
    \toprule \diagbox [width=15em,trim=l] {Data}{Value and error} & $\delta_{1}$
    & $\delta_{2}$ & error of $\delta_{1}$ & error of $\delta_{2}$ \\ \hline
    Correct data & 0.5 & 1 & 0 & 0 \\
    Clean training data & 0.49565 & 0.98456 & $0.8\%$ & $1.5\%$ \\
    Training data with $1\%$ noise & 0.50601 & 0.97770 & $1.2\%$ & $2.2\%$ \\
    \bottomrule
    \end{tabular}
    \vspace{0cm}
\end{table*}

%%%%%%%%%%%%%%%%%%%%%%%%%%%%%%%%%%%%%%%%%%%%%%%%%%%%%%%%%%%%%%%%%%%%%%%%%%%%%%%%%

\section{Conclusions}\label{sec5}

We have here put forward the PINN-based method to predict nonlinear dynamics of necklace patterns carrying integer, half-integer, or non-integer reduced angular momentum (i.e., the total momentum divided by the field's total power) in the framework of the NLSE (nonlinear Schr\"{o}dinger equation) and GPE (Gross-Pitaevskii equation), which are fundamental equations for nonlinear optics and Bose-Einstein condensates. The results show that the PINN method has the ability to characterize and predict the propagation of the necklace patterns. The contraction or expansion of the necklace structure is precisely predicted. The necklaces carrying different values of the angular momentum, built as superpositions of individual quasi-solitons, exhibit effectively stable propagation, while the individual constituents are unstable (as the TSs (Townes solitons)), if considered in isolation. The specific nonlinear dynamics of the necklace beams are accurately predicted by the PINN, which, in turn, indicates the efficiency and accuracy of the method. We have also explored effects of different items on the performance of PINN. The inverse problem of the data-driven discovery of the model's parameters has been addressed too, revealing the potential advantage of using PINN for solving high-dimensional inverse problems.

Assets of the PINN suggest a possibility to learn physical constraints corresponding to multiple different scenarios in broader settings, thus resulting in a robust generalization for the multi-scenario modeling. Furthermore, we expect that nonlinear dynamical properties of various species of spatial optical solitons can also be revealed by the PINN deep neural network. In particular, the self-trapped necklace beams, similar to those studied in detail in this work, can be constructed in many
nonlinear systems. It is also possible to develop similar methods to address the evolution dynamics of necklace-like matter-wave solitons and quantum droplets in Bose-Einstein condensates. In addition to that, a challenging possibility is to explore the dynamics of 2D necklace-shaped modes driven by the GPE written for curved surfaces \cite{craps_maximally_2017,tononi_bose-einstein_2019}, the simplest example given by Eq. (\ref{conus}).

\begin{backmatter}
    \bmsection{Funding}
    This work was partially supported by the National Natural Science Foundation of China (No. 61975130), and Guangdong Basic and Applied Basic Research Foundation (No. 2021A1515010084). The work of B.A.M. is supported, in part, by the Israel Science Foundation through grant No. 1695/22.

    \bmsection{Disclosures}
    The authors declare no conflicts of interest.

    \bmsection{Data availability}
    Data underlying the results presented in this paper are not publicly available at this time but may be obtained from the authors upon reasonable request.
\end{backmatter}

% \appendix

% \section{Section title of first appendix\label{app1}}

% This Appendix presents key codes for all algorithms employed in the paper. The underlying data for the results reported in this paper are not included here, but may be obtained from the authors upon a reasonable request.

% In the structure of the PINN, the complex-valued $\widehat{\psi}(z,x,y)=[\widehat{u}(z,x,y),\widehat{v}(z,x,y)]$ can be defined in Python as
% \begin{lstlisting}
% def psi(z, x, y):
%     psi = neural_net(tf.concat([z, x, y],
%           1), weights, biases)
%     u = psi[:, :, 0:1]
%     v = psi[:, :, 1:2]
%     return u,v
% \end{lstlisting}

% Next, the residual PINN $f(z,x,y)$ is implemented as

% \begin{lstlisting}
% def f_uv(z, x, y):
%     u, v = self.psi(z, x, y)
%     u_x = tf.gradients(u, x)[0]
%     v_x = tf.gradients(v, x)[0]
%     u_y = tf.gradients(u, y)[0]
%     v_y = tf.gradients(v, y)[0]
%     u_z = tf.gradients(u, z)[0]
%     v_z = tf.gradients(v, z)[0]
%     u_xx = tf.gradients(u_x, x)[0]
%     u_yy = tf.gradients(u_y, y)[0]
%     v_xx = tf.gradients(v_x, x)[0]
%     v_yy = tf.gradients(v_y, y)[0]
%     f_re = v_z - 0.5 * u_xx - 0.5 * u_yy - (u ** 2 + v ** 2) * u
%     f_im = u_z + 0.5 * v_xx + 0.5 * v_yy + (u ** 2 + v ** 2) * v  
%     return f_re,f_im
% \end{lstlisting}

% \bibliographystyle{opticajnl}
% \bibliography{reference}

\end{document}